\documentclass[journal]{IEEEtai}
\usepackage[colorlinks,urlcolor=blue,linkcolor=blue,citecolor=blue]{hyperref}
\usepackage{color,array}
\usepackage{graphicx}
\usepackage{amsmath}
\begin{document}

\title{Decentralized Deep Learning for Multi-Access Edge Computing: A Survey on Communication Efficiency and Trustworthiness}

\author{Yuwei Sun, Hideya Ochiai, and Hiroshi Esaki, \IEEEmembership{Member, IEEE}
\thanks{19 November 2021. This work was supported in part by the JRA program at the RIKEN Center for Advanced Intelligence Project.}
\thanks{Yuwei Sun, Hideya Ochiai, and Hiroshi Esaki are with the Graduate School of Information Science and Technology, University of Tokyo, Tokyo, 1138654 Japan (e-mail: ywsun@g.ecc.u-tokyo.ac.jp, ochiai@elab.ic.i.u-tokyo.ac.jp, hiroshi@wide.ad.jp).}
\thanks{This paragraph will include the Associate Editor who handled your paper.}}

\markboth{Journal of IEEE Transactions on Artificial Intelligence, Vol. 00, No. 0, Month 2020}
{Yuwei Sun \MakeLowercase{\textit{et al.}}: Decentralized Deep Learning for Multi-Access Edge Computing: A Survey}

\maketitle

\begin{abstract}
Wider coverage and a better solution to a latency reduction in 5G necessitate its combination with multi-access edge computing (MEC) technology. Decentralized deep learning (DDL) such as federated learning and swarm learning as a promising solution to privacy-preserving data processing for millions of smart edge devices, leverages distributed computing of multi-layer neural networks within the networking of local clients, whereas, without disclosing the original local training data. Notably, in industries such as finance and healthcare where sensitive data of transactions and personal medical records is cautiously maintained, DDL can facilitate the collaboration among these institutes to improve the performance of trained models while protecting the data privacy of participating clients. In this survey paper, we demonstrate the technical fundamentals of DDL that benefit many walks of society through decentralized learning. Furthermore, we offer a comprehensive overview of the current state-of-the-art in the field by outlining the challenges of DDL and the most relevant solutions from novel perspectives of communication efficiency and trustworthiness.
\end{abstract}

\begin{IEEEImpStatement}
The proliferation of smart devices and edge applications based on deep learning is reshaping the contours of future high-performance edge computing, such as intelligent environment sensors, autonomous vehicles, smart grids, and so forth. Decentralized deep learning (DDL) as a key enabler of the Multi-access Edge Computing benefits society through distributed model training and globally shared training knowledge. However, crucial fundamental challenges have to be overcome in the first place to make DDL feasible and scalable, which are decentralization techniques, communication efficiency, and trustworthiness. This survey offers a comprehensive overview from the above perspectives, suggesting that DDL is being intensively studied, especially in terms of privacy protection, edge heterogeneity, and adversarial attacks and defenses. Moreover, the future trends of DDL put weight on topics like efficient resource allocation, asynchronous communication, and fully decentralized frameworks.
\end{IEEEImpStatement}


\section{Introduction}
Deep learning (DL) was first proposed to solve problems where a set of training data was collected for centralized data processing. In recent years, with the rapid advancement in this field, its applications have extended to various industries, benefiting people’s life. However, collecting and transmitting such enormous data into centralized storage facilities is usually time-consuming, inefficient, and with privacy concerns. Limitations in network bandwidths and so on could bring in high latency. Moreover, the risk of personal data breaches correlated with data transmission to a centralized computing recourse causes data privacy concerns. Especially, with the increase of data privacy awareness in society,  legal restrictions such as the General Data Protection Regulation (GDPR) \cite{1} have been promoted making such a centralized framework even unpractical.    
 
In contrast, compared with a centralized framework where clients have to provide raw data to a central server for the model training, in a decentralized framework, the sensitive data of a client is processed directly on its local device. The concept of decentralized deep learning (DDL) was first proposed to facilitate the training of a deep network with billions of parameters using tens of thousands of CPU cores \cite{2}. A few years later, the famous federated learning (FL) was proposed by Google \cite{3}, allowing privacy-preserving collaborative learning among edge devices by leveraging on-device model training and trained model sharing. For one thing, local model training greatly reduces the latency in a centralized framework. Another important point is that a large system consisting of thousands of clients improves its performance by aggregating the results from local model training, without disclosing raw training data. 

Despite the success of FL, in real life, a participating local device typically necessitates certain qualifications for efficient local model training. Limitations in device memory and computation capability can greatly increase the local training time of a client, and network bandwidth limitations can result in the increase of clients’ waiting time for transferring models thus causing a delay in an FL training cycle. Furthermore, non-independent and identically distributed (Non-IID) data of clients results in time-consuming convergence of FL. To tackle the challenge of communication efficiency in DDL approaches such as split learning (SL) and smart client selection have been proposed.

\begin{figure}
\centerline{\includegraphics[width=\linewidth]{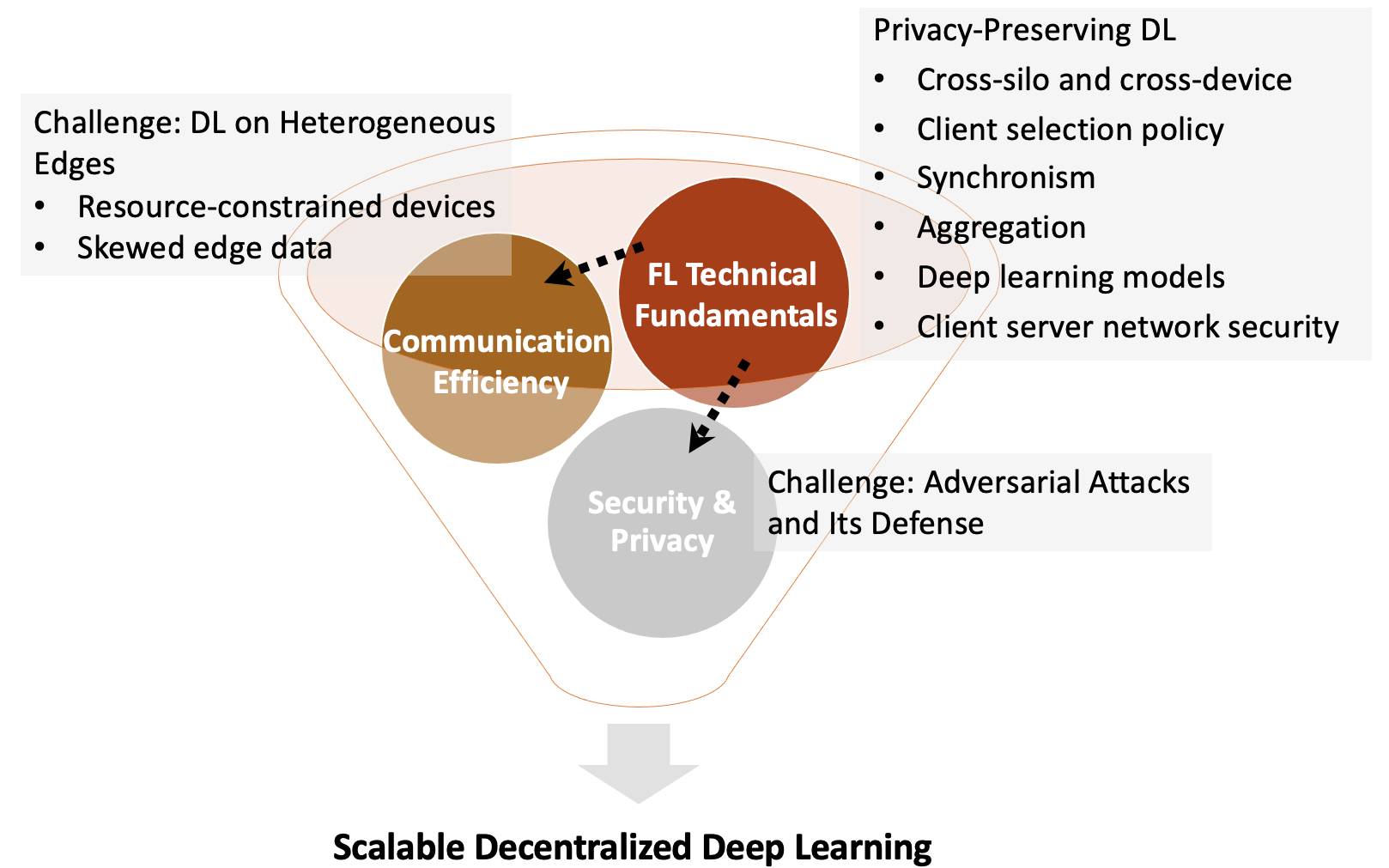}}
  \caption{Three pillars for scalable decentralized deep learning.}
  \label{sun1}
\end{figure}

Moreover, towards a future integrated society by leveraging multi-agent multi-access edge computing, it necessitates building trust in such emerging technologies, i.e. trustworthiness. Nevertheless, recent works have demonstrated that FL may not always provide sufficient guarantees to personal data privacy and deep learning model integrity. Even in a decentralized framework like FL, an attacker still can compromise systems by injecting a trojan into either a client's local training data or its local model, and such an attack can further expand its influence to other clients through model sharing. In other cases, an attacker could even steal information from clients by observing the transmitted model gradients. To overcome these threats, defense strategies aiming to improve systems robustness and detect malicious behaviors are applied in FL. To this end, there are three pillars for the development of scalable decentralized deep learning covering FL technical fundamentals, communication efficiency, and security and privacy (trustworthiness) (Fig. \ref{sun1}).
 
This survey paper is organized as follows. Section \ref{sec2} comprehensively demonstrates the technical fundamentals for facilitating DDL and relevant applications in various fields. Section \ref{sec3} presents the main challenges and promising methodologies for future scalable DDL, from the perspectives of communication efficiency under edge heterogeneity and trustworthiness. Section \ref{sec4} concludes the paper, discussing open challenges and future directions.

\section{Towards Decentralized Deep Learning}
\label{sec2}
\subsection{Multi-Access Edge Computing}
According to an annual report from Nokia in 2020 \cite{4}, a large increase in the number of broadband IoT and Critical IoT devices will be observed in the next five years, such as AR, VR, and cloud robotics. Likewise, the number of massive IoT devices like different types of meters and sensors is to increase greatly as well. Most of these devices will be operated based on Artificial Intelligence (AI). In this regard, with the proliferation of smart devices and applications at the network edge based on AI such as intelligent environment sensors, autonomous vehicles, health care, smart grid, and so on, AI is playing a key role in data processing, knowledge acquisition, and resource management. For instance, AI has been used in edge service optimization in the Internet of Vehicles (IoV) \cite{xu2022} and other compelling applications for collective intelligence in wireless networks \cite{WangJZRCH20}. Traditionally, data generated on a smart device is sent to a remote computing server for processing. Though 5G aims to provide greater connectivity for multi-type devices with a big boost in the speed of handling big data, there still needs a wider coverage to facilitate efficient data processing. For this reason, a better solution to latency reduction is to combine with multi-access edge computing (MEC) technology. The inextricably correlated MEC reduces latency by leveraging compute resources in a network closer to the end-users, e.g., a local server and the gateway.

\subsection{Data Privacy and Decentralized Deep Learning }
The mathematical model of the perceptron was first proposed back in 1958 \cite{5}, which is a probabilistic model for information storage and organization. The multi-layer perceptron \cite{6} adds to the practicability of neural networks, as a useful alternative to traditional statistical modeling techniques. Lecun et al. \cite{7} presented deep learning (DL) that allows computational models composed of multiple processing layers to learn representations of data with multiple levels of abstraction. Nowadays, various DL models have been developed and broadly adopted in many walks of society, such as convolutional neural networks (CNNs), recurrent neural networks (RNNs), and so forth. 

\begin{figure*}
\centerline{\includegraphics[width=0.78\linewidth]{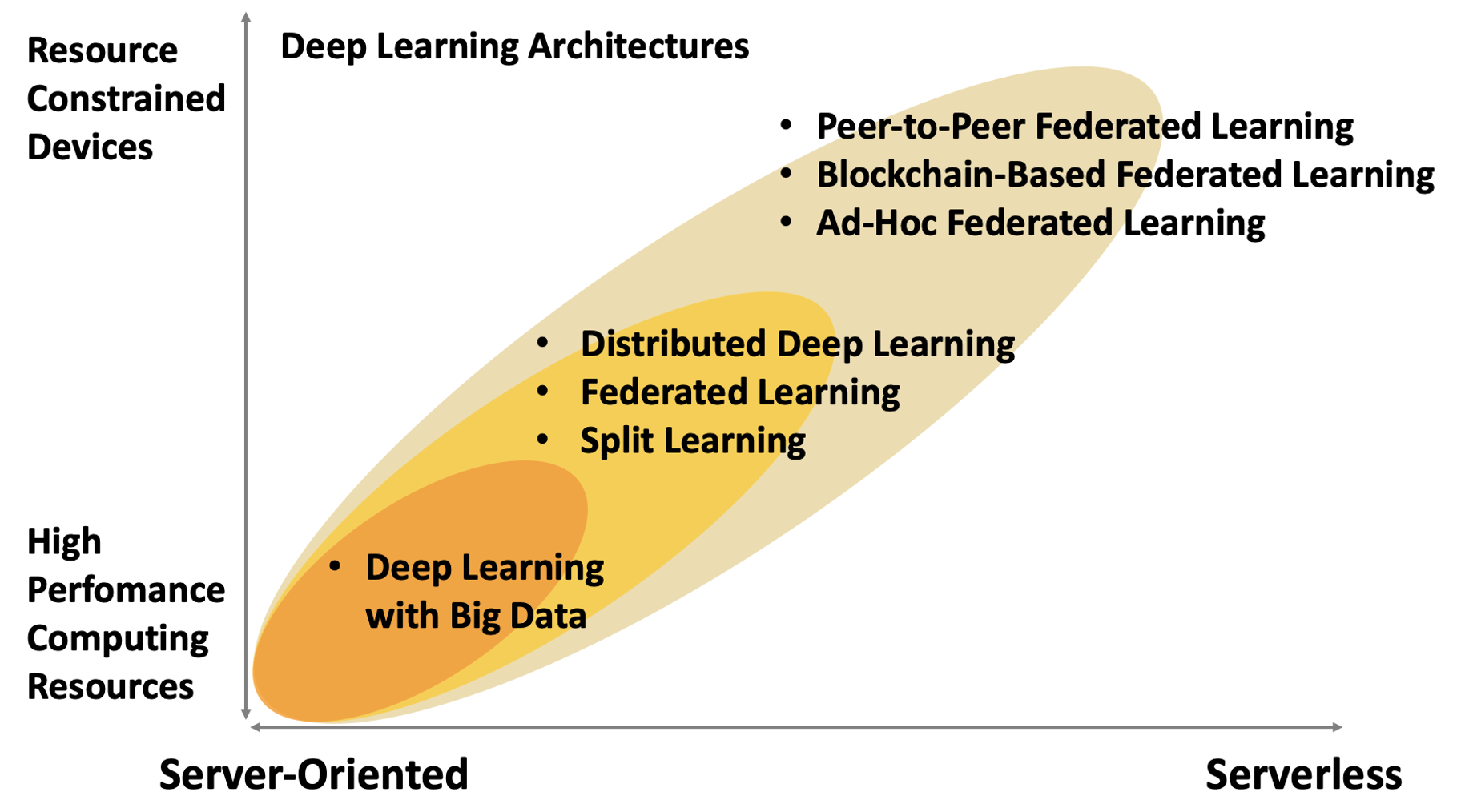}}
  \caption{The rising of decentralized deep learning.}
  \label{sun2}
\end{figure*}

Moreover, there are mainly two topologies of DL for processing distributed data, i.e. centralized DL (server-oriented) and decentralized DL (client-oriented or server-less) (Fig. \ref{sun2}). A centralized or stand-alone framework leverages a central high-performance computing resource to achieve the desired model performance by collecting data from various data sources. In this case, the collected data is usually exposed to the AI algorithm on the cloud. In contrast, a decentralized framework is considered as a privacy-preserving architecture by leveraging local model training based on distributed data sources on resource-constrained devices like smartphones. Since its introduction, the decentralized framework \cite{dml} has proliferated in academia and industry. Li et al.\cite{muli} further extended the concept of the parameter server framework and demonstrated a robust, versatile, and high-performance implementation, capable of handling a diverse array of algorithms for distributed machine learning problems based on local training data. Moreover, in recent years, federated learning (FL) has become one of the most famous decentralized frameworks, which was proposed by Google initially to improve the Google Keyboard (Gboard)’s performance in next word prediction \cite{3}. The architecture of FL allows users to take full advantage of an AI algorithm without disclosing their original local training data, bridging the gap between centralized computing resources and distributed data sources. FL achieves a better model by leveraging globally shared model parameters. 

Furthermore, a fully decentralized framework refers to server-less architectures based on technologies such as the blockchain and edge consensus \cite{sl, 52, 54, 61, 66, kim20}, and ad hoc network \cite{adhoc}. For instance, Swarm Learning (SL) \cite{sl} is a decentralized approach that combines edge computing, blockchain-based peer-to-peer networking, and other state-of-the-art decentralization technologies for classifying diseases with distributed medical data. Moreover, Li et al. \cite{52} presented a decentralized federated learning framework based on blockchain for the global model storage and the local model update exchange, where the local updates are encrypted and stored in blocks of the blockchain after the consensus by the committee. Similarly, Kim et al.\cite{kim20} demonstrated an end-to-end latency model of chained federated learning architecture, with the optimal block generation rate decided by communication, computation, and consensus delays.

In recent years, data privacy has been a major concern, exacerbated by social events such as the Cambridge Analytica scandal \cite{8} and FBI-Apple encryption dispute \cite{9}. Data privacy concerns associated with the centralized data processing of a traditional DL pipeline necessitate more considerations on privacy-preserving system design and data protection strategies. To this end, the decentralized framework provides a promising solution to data privacy in large-scale multi-agent collaborative learning. For instance, massively decentralized nodes can be applied to diverse use cases, such as industrial IoT \cite{10}, environment monitoring using diverse sensors \cite{11}, human behavior recognition from surveillance cameras \cite{12}, robotics, and connected autonomous vehicles control \cite{13, 14}, federated network intrusion detection across multiple parties \cite{15, 16} and so forth.

\subsection{Federated Learning from a Network System Perspective}
\subsubsection{Cross-silo and Cross-device}
\begin{figure}
\centerline{\includegraphics[width=\linewidth, trim=1 1 1 1,clip]{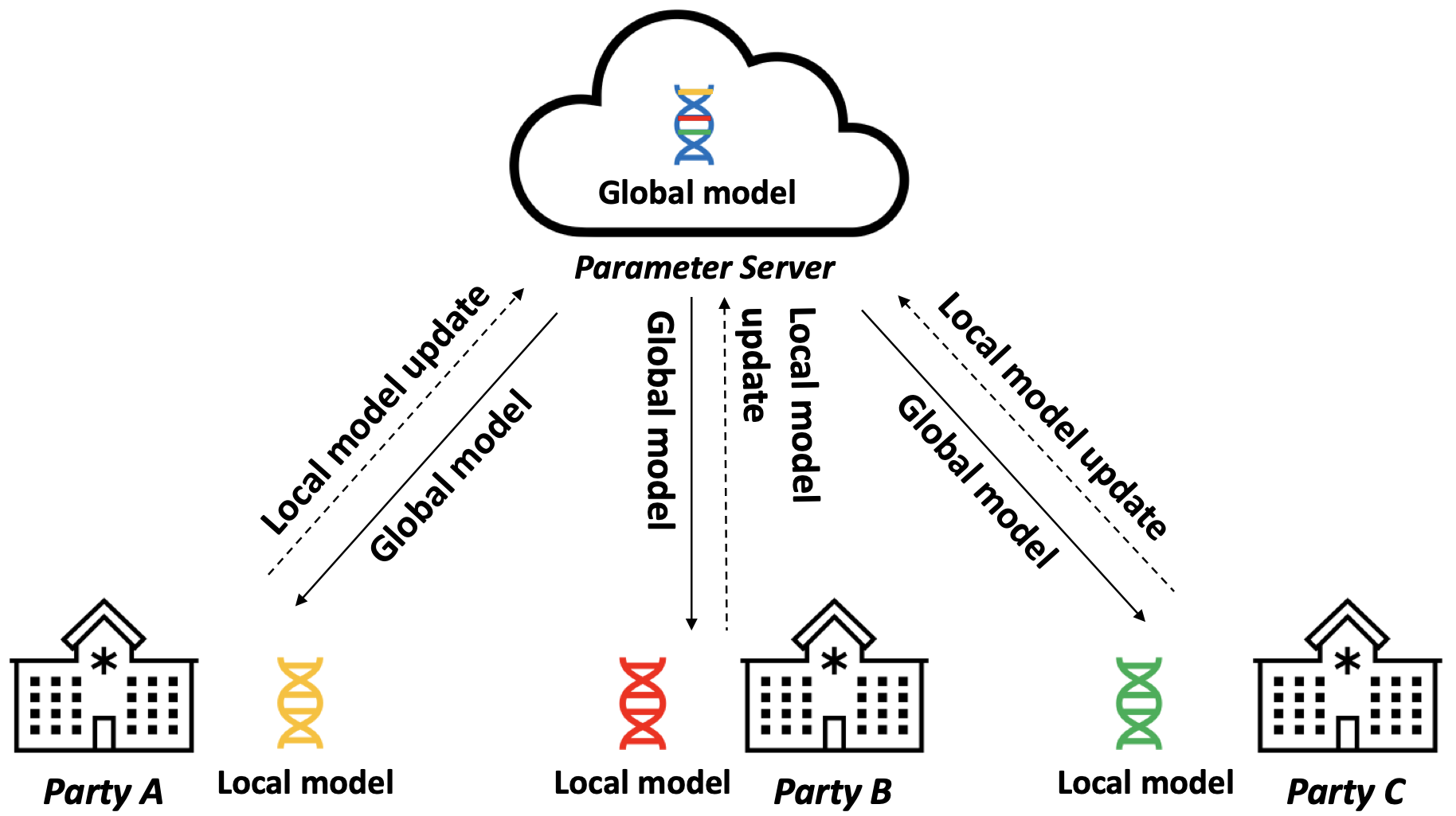}}
  \caption{Federated learning leverages multi-party model sharing sharing for privacy-preserving machine learning.}
  \label{sun3}
\end{figure}

The cross-silo setting of FL represents a scenario of multi-party collaborative model training (Fig. \ref{sun3}), where collected data from local devices is sent to an edge server located inside the organization for computation. In this case, an upper-level remote server is applied for further computation. Data is transparent to all clients inside the organization, but it would not be exposed outside it. For instance, healthcare institutes could adopt this scheme to share medical images for identifying a rare disease \cite{18}. In this case, the cross-silo setting allows the institutes to share insights on the disease under data protection. On the other hand, a cross-device setting is a more rigorous scenario that collected data should not leave a device. It necessitates efficient on-device computing and timely model transmission to a remote server directly.    

\subsubsection{Client Selection Policy}
Typically, to reduce the latency of waiting, at each round, the Parameter Server (PS) randomly selects only a small subset k out of m clients for the local model training and broadcasts the current global model w to the selected clients. Then, starting from w, each client$i$ updates its local model wi by training on its data and transmits the local model update back to the PS. In addition, other client selection policies such as cluster-based selection and reinforcement learning-based selection are adopted to reduce the time cost for global model convergence (see \ref{Section 3.1.2}).

\subsubsection{Synchronism}
There are two types of client scheduling approaches, i.e., synchronous FL and asynchronous FL. In synchronous FL, at each round, the PS waits for the completion of all allocated local training. However, in this case, the slowest local training task due to a relatively large data volume, computing device constraints, and so on becomes the bottleneck of training. In contrast, the asynchronous FL allows a client to upload the local update at any stage of the training. Besides, a client can offer multiple functions including local model training, network traffic transit, and so forth \cite{19, 69}.    

\subsubsection{Aggregation} 
As aforementioned, the next round’s global model takes the value of all local model updates’ aggregation results. Federated averaging (FedAvg) \cite{17} computes a weighted average as in (\ref{eq1}) to update the global model, given the volume of local training data is varying from client to client (the contribution is varying). 

\begin{equation}
\label{eq1} 
w_{t+1} = w_t + \sum_{i\in k}\frac{n_i}{n_k}(w_t^i - w_t)
\end{equation}

Where $w_t$ represents the weights of the current global model, $w_{t+1}$ is the weights of the next round’s updated global model, $w_t^i$ is the weights of client$i$'s trained local model, and $n_i$ and $n_k$ represent respectively the volume of client$i$'s local training data and that of the total training data from all the selected clients. 

Moreover, robust aggregation strategies aim to drop a malicious update by measuring similarity among local model updates (see \ref{Section 3.2.5}). According to a local update's integrity, only qualified updates are aggregated into the global model at each round.   

\subsubsection{Deep Learning Models}  
Most of the current studies and applications around FL are based on a supervised model, where the model is trained on labeled data for typically a classification task. However, in real life, collected data is usually unlabeled and a supervised model is not compatible. Deep learning models including unsupervised learning and reinforcement learning are not sufficiently studied in the context of FL. For instance, by leveraging FL for a reinforcement task of robotics, a global agent could learn multiple action policies efficiently from diverse environments at the same time \cite{20}. 

\subsubsection{Client Server Network Security}
From the perspective of a network system, FL encounters threats from mainly three components of the systems, i.e., the parameter server (PS), the client, and the transmission pathway. The PS is usually well secured and highly maintained compared with edge devices. Besides, the communication between the PS and a client is also commonly protected through end-to-end encryption. On the other hand, though the integrity of a client is verified to participate in the FL training, an edge still encounters intrusion by an adversary due to its relatively incomplete defense strategies taken at local. In FL, due to all clients have equal access to the global model through the aggregated model broadcast at each round, it provides a huge attacking surface for the adversary to compromise the systems. To this end, we consider that a compromised edge is the main threat to FL systems (see also \ref{Section 3.2}).
  
\section{Challenges and Methodologies towards Scalable Decentralized Deep Learning}
\label{sec3}
\subsection{Communication Efficiency Under Edge Heterogeneity} 
Communication efficiency is an important contributor to evaluating the performance and scalability of distributed processing. Decentralized deep learning (DDL) can reduce computation time by synchronizing the different models during training. However, this leads to an increase in communication cost as the model size increases or the convergence becomes slow. Notably, one of the largest challenges of scaling FL in real life today is that device qualifications at the edge are varying. In particular, such heterogeneity lies in two main aspects, i.e., device capability and data distribution. 

Firstly, for the device capability, especially in the case of cross-device FL, a DL model is usually operated on a resource-constrained mobile device such as smartphones. The capabilities of these mobile devices are varying due to hardware limitations and cost constraints. Moreover, the network bandwidth of a local area network (LAN) also greatly limits the model transmission efficiency, resulting in a delay in the decentralized learning cycle. 

Secondly, for the data distribution, samples held by different clients are typically diverse with different data sizes and distributions, i.e., non-independent and identical distributed (non-IID). For example, in a multi-classification task with $C$ categories. Each Client$k$ owns a local dataset $D^{(k)}$ that consists of samples with unbalanced labels $\{1,2,…,C\}$. Then, Client1 has 80\% samples from Label1 and Client2 has 80\% samples from Label2. Mathematically speaking, suppose that $f_w:x\rightarrow y$ denotes a supervised neural network classifier with parameters $w$, taking an input $x_i \in x$ and outputting a real-valued vector where the $j$th element of the output vector represents the probability that $x_i$ is recognized as class $j$. Given $f_w(x)$, the prediction is given by $\hat{y} = \mbox{arg}\max_j f_w(x)_j$ where $f_w(x)_j$ denotes the $j$th element of $f_w(x)$. We assume the common data distribution $p(x|y)$ is shared by all clients in FL, and client$i$ has $p_i(y)$. Then when samples held by clients are skewed with various $p_i(y)$, $p_i(x,y)=p(x|y)p_i(y)\,s.t.\,p_i(y)\neq p_j(y)$, for all $i\neq j$. Client1 follows $p_1(x,y)$ and Client2 follows $p_2(x,y)$, i.e., they are non-IID. Though the random client selection policy in classical FL aims to reduce the time for waiting, the non-IID local data of the selected clients could give rise to a time-consuming convergence or even failing to converge the global model. In this section, we demonstrate the most relevant methodologies used to spread and reduce the amount of data exchanged between the server and clients tackling the edge heterogeneity problem (Table \ref{tab:freq}).     

\begin{table*}
\centering
  \caption{Methodologies for Improving communication efficiency under Edge Heterogeneity}
  \label{tab:freq}
  \footnotesize
  \begin{tabular}{lllll}

    Challenge & Work & Year &Methodology&Application\\

    Resource-Constrained 		   & Vepakomma et al. \cite{22} & 2018 & Split Learning& Image classification\\
    Edge   									     & Nishio et al. \cite{FedCS} & 2018 & Resource scheduling  &  Image classification\\   
    												  & Singh et al. \cite{23} & 2019 & Split Learning& IoT, Healthcare\\
    												  & Thapa et al. \cite{21} & 2020 & Split Learning& Healthcare, Image classification \\
    												  & Khan et al. \cite{19} & 2020 & Stackelberg game theory& Image classification\\

  Data Heterogeneity & Jeong et al. \cite{FAug} & 2018 & Federated Augmentation & Image classification\\
    									& Sener et al. \cite{25} & 2018 & K-Center clustering & Image classification\\
   	 									& Zhao et al. \cite{noniid} & 2018 & Data-sharing strategy & Image classification\\
     									& Wang et al. \cite{26} & 2020 & Reinforcement Learning & Image classification\\
                                        & Duan et al. \cite{astraea} & 2020 & Data augmentation and rescheduling & Image classification\\  
     									& Sun et al. \cite{13} & 2021 & Segmented Federated Learning & Cybersecurity\\

\end{tabular}
\footnotesize
\end{table*}

\subsubsection{Resource-Constrained Edge}
\begin{figure}
\centerline{\includegraphics[width=\linewidth, trim=1 1 1 1,clip]{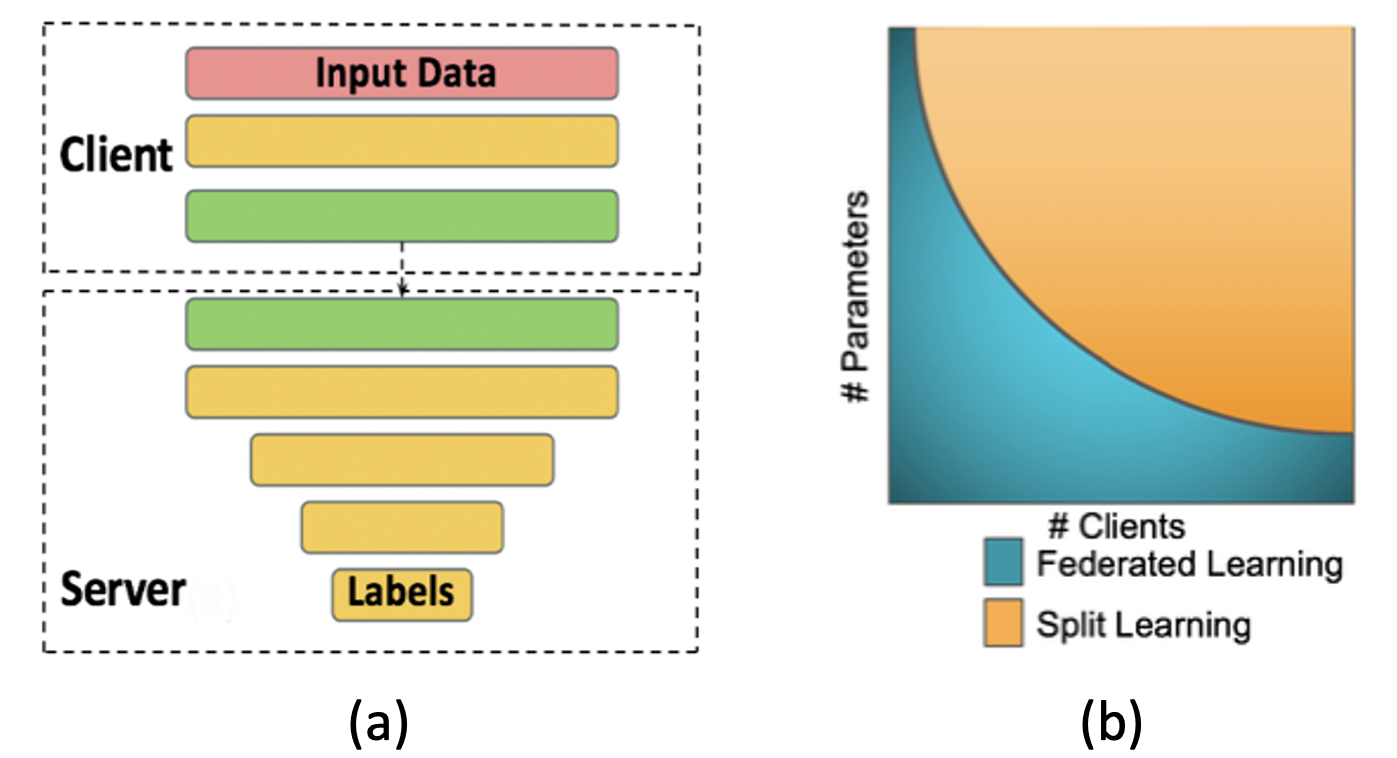}}
  \caption{(a) The architecture of the vanilla Split Learning \cite{22}. (b) A model performance comparison between FL and SL regarding the number of total model parameters and the number of total clients. The hyperbola shows the regions where one model outperforms the other regarding communication efficiency \cite{23}.}
  \label{sun4}
\end{figure}

Despite FL allowing each client to train its local model, the communication efficiency of FL is largely limited by the client-side qualifications such as network bandwidth, device memory, and computation capability, and so on. Under these circumstances, Split Learning (SL) \cite{21, 22} was proposed to facilitate model training based on edge cloud computing. In SL, a complicated DL model is partitioned into two sub-networks based on a specific layer called the cut layer, and then these sub-networks are trained on the client and the PS respectively (Fig. \ref{sun4}(a)). For each round, the client leverages forward propagation of its local sub-network based on local data, and then sends the intermediate representation of local data at the cut layer together with labels (vanilla Split Learning) to the PS for completing the forward propagation and the computation of loss. Finally, the gradients of the cloud-side sub-network are computed using back propagation, and they are sent back to the client for updating its local model. As such, for each round’s training, several times of transmission between the client and the PS are necessary. 

Moreover, a comprehensive comparison of communication efficiency between SL and FL was presented by Singh et al. \cite{23}. To study the relationship between communication efficiency and factors such as the total client number and model parameter number, a trade-off between the two models was demonstrated (Fig. \ref{sun4}(b)), where the hyperbola shows the regions where one model outperforms the other regarding communication efficiency, in other words, less data transmission between the client and the PS. Besides, by comparing SL and FL in real-life scenarios of smart watches with users in a diverse range from hundreds to millions, the result suggests that SL is more efficient and scalable when it comes to a relatively large number of clients and a relatively large DL model.   

Furthermore, to tackle various constraints of computational resources and wireless channel conditions, Nishio et al. \cite{FedCS} demonstrated a method called FedCS. In FedCS, the PS estimates the time required for conducting several steps of FL based on resource information of clients and schedules clients such that it aggregates as many client updates as possible to accelerate performance improvement during training. It shows a significantly shorter training time compared with the classical FL. In addition, Khan et al. \cite{19} proposed an incentive-based FL framework based on the Stackelberg game theory to motivate the participation of devices in the learning process, while optimizing the client selection for minimizing the overall training cost of computation and communication.

\subsubsection{Data Heterogeneity} 
\label{Section 3.1.2}
\begin{figure}
\centerline{\includegraphics[width=0.8\linewidth, trim=1 1 1 1,clip]{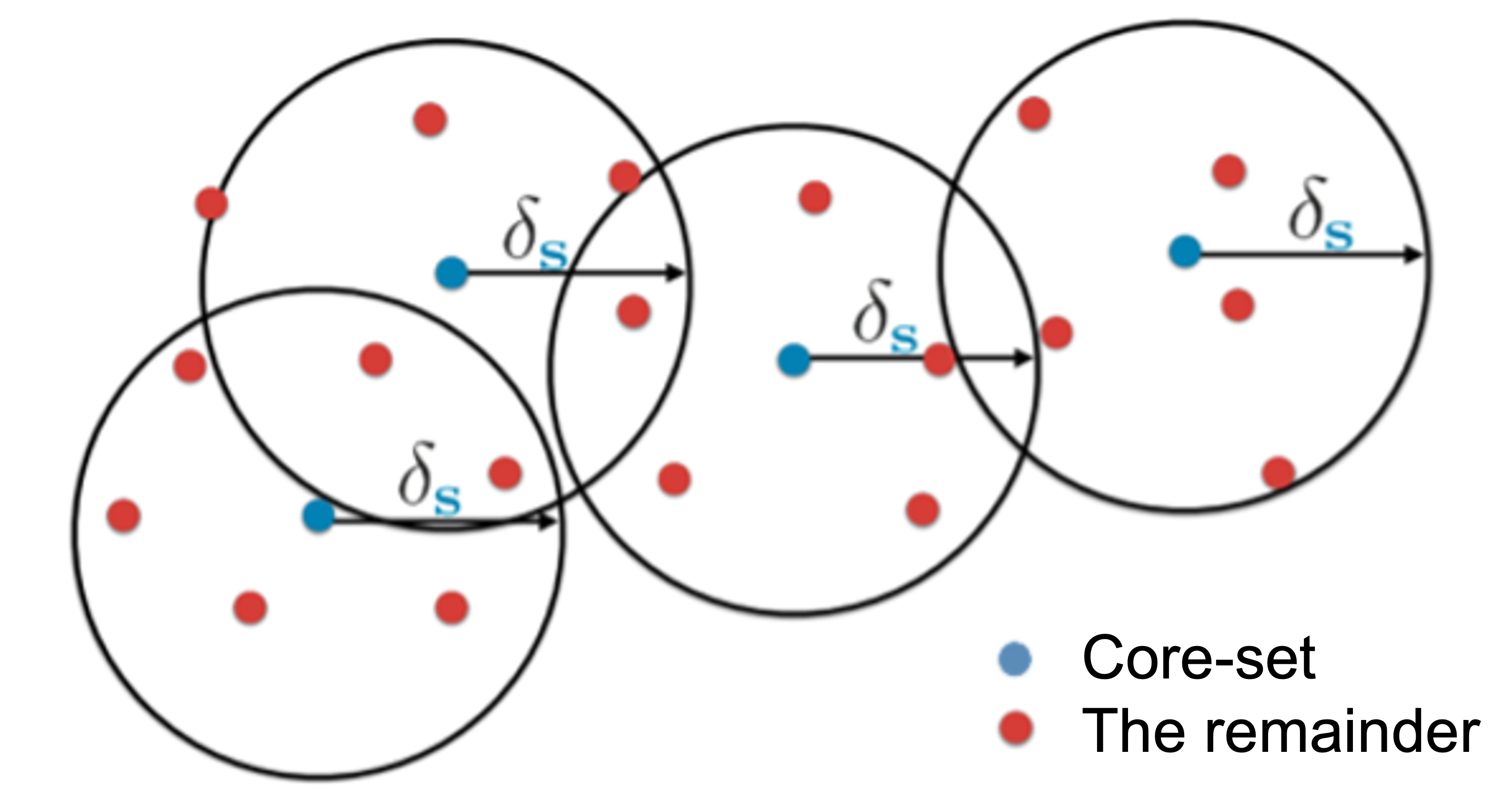}}
  \caption{K-Center clustering algorithm aims to find a core-set (blue points) to represent the whole dataset (blue and red points) when conducting the training. \cite{25}}
  \label{sun5}
\end{figure}

Though the random client selection in FedAvg has been working well given data samples held by different clients are independent and identical decentralized (IID) \cite{24}. Unfortunately, this scheme doesn’t work well when applied to real-world data samples, which are typically non-IID. For this reason, an efficient client selection policy during FL training is critical for the fast convergence of the global model, instead of the random client selection policy. Sener et al. \cite{25} presented the K-Center clustering algorithm for choosing images to be adopted from a very large collection. They aim to find a subset such that the performance of the model on the labeled subset and that on the whole dataset will be as close as possible (Fig. \ref{sun5}). Furthermore, by leveraging the K-Center algorithm in FL under the non-IID settings, participating clients can be clustered into various groups based on their data distributions. Then, by carefully selecting clients from each group during training, it contributes to a faster global model convergence and performance improvement \cite{26}.  

\begin{figure}
\centerline{\includegraphics[width=0.8\linewidth]{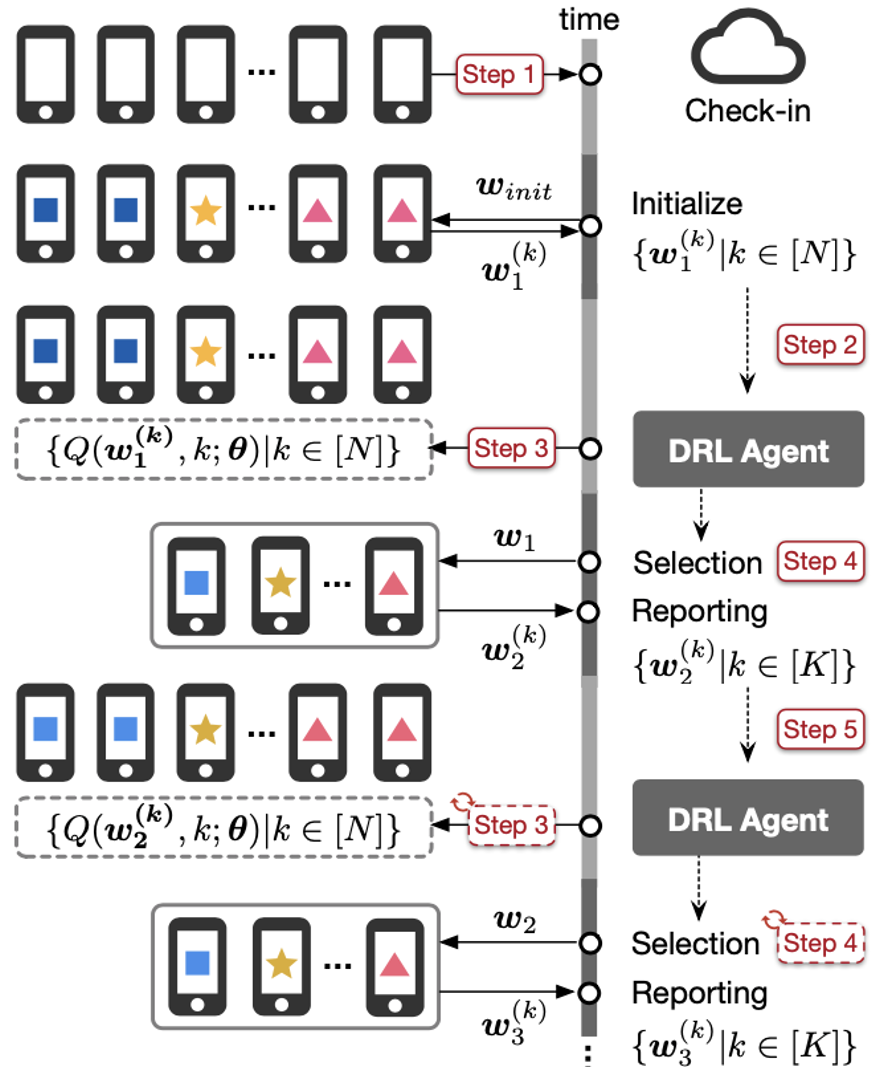}}
  \caption{The RL-based client selection policy for faster convergence of FL. \cite{26}}
 \label{sun6}
\end{figure}

Similarly, a reinforcement learning (RL)-based FL on non-IID data was presented by Wang et al \cite{26}, where an experience-driven control framework called FAVOR intelligently chooses the clients to participate in each round of FL (Fig. \ref{sun6}). This approach aims to counterbalance the bias introduced by non-IID data, thus speeding up the global model convergence.  The objective of this approach is to achieve the desired model performance within the fewest rounds. In particular, deep Q-learning (DQN) is adopted to learn how to select a subset of clients at each round thus maximizing a reward computed from the current reward and expected future rewards. Besides, there are three main components w.r.t. RL, i.e., the state, the action, and the reward \cite{27}. Here the state of the environment is defined as compressed weights of the global model and local models. The available actions for the RL agent are a large space of size $\binom{N}{K} $, where $K$ is the total number of clients and $N$ is the number of selected clients at each round of FL. Finally, the reward of the DQN agent consists of the incentive from achieving high accuracy and the penalty for taking more rounds to achieve the desired performance. 
   
Moreover, Segmented Federated Learning (Segmented-FL) was proposed to tackle data heterogeneity in network intrusion detection \cite{13}. Participants with highly skewed non-IID network traffic data are separated into various groups for personalized federated learning based on their recent behavior. Then, each group is assigned an individual global model for the aggregation respectively. Besides, for each round's training, a new group segmentation is formed and the global model of a group is updated based on the weighted averaging of its current global model, local model updates from the group, and the other existing global models. Consequently, it shows that Segmented-FL for network intrusion detection with massively distributed data sources outperforms the classical FL. 

In addition, Zhao et al. \cite{noniid} presented a data-sharing strategy to improve training on non-IID data by creating a small data subset that is globally shared between all the edge devices. The experiments show that accuracy can be increased by ~30\% for the CIFAR10 dataset with only 5\% globally shared data, compared with the accuracy of FL. Likewise, Jeong et al. \cite{FAug} proposed federated augmentation (FAug) where each device trains a generative model, and thereby augments its local data towards yielding an IID dataset. The result shows around 26x less communication overhead for achieving the desired test accuracy compared to FL. 

\subsection{Trustworthiness}
\label{Section 3.2}
A decentralized framework such as federated learning (FL) encounters threats from malicious AI. As aforementioned in Section 2, threats from an edge client are more common and critical for FL, compared with the security of the server and the middle data transmission. FL extends the surface for the attacker to compromise one or several participants. In this regard, an adversary can intrude such decentralized systems through a compromised edge as a backdoor, by either manipulating local training data \cite{28, 29, 30, 31} or replacing a local model update \cite{28, 31, 32}, triggering attacker-desired behavior. This kind of attack extends its influence to other clients in the systems through malicious model sharing with poisoned model weights. (see \ref{Section 3.2.1}, \ref{Section 3.2.2}, \ref{Section 3.2.3}, \ref{Section 3.2.4})

In the contrast, the controversy surrounding threats on FL has also been intensively discussed in recent years. These defense strategies can mainly be separated into two categories, i.e., robust aggregation and anomaly detection. For the robust aggregation, it is related to improving the resilience of an aggregation algorithm (e.g. FedAvg) by either carefully selecting the local models for aggregation \cite{33, 34, 35} or adding noise to the aggregated model for counterbalancing the malicious update\cite{31, 37, 38}. On the other hand, various anomaly detection approaches are leveraged for identifying a malicious local model update, including comparing the similarity between local updates and finding the ones greatly diverging from the others \cite{39, 40, 41, 42, 64}, applying a cloud validation set with a small number of data samples from each client \cite{61}, and so on. (see \ref{Section 3.2.5})

\subsubsection{Threat Models}
\label{Section 3.2.1}
\begin{figure*}
\centerline{\includegraphics[width=0.8\linewidth, trim=2 2 2 2,clip]{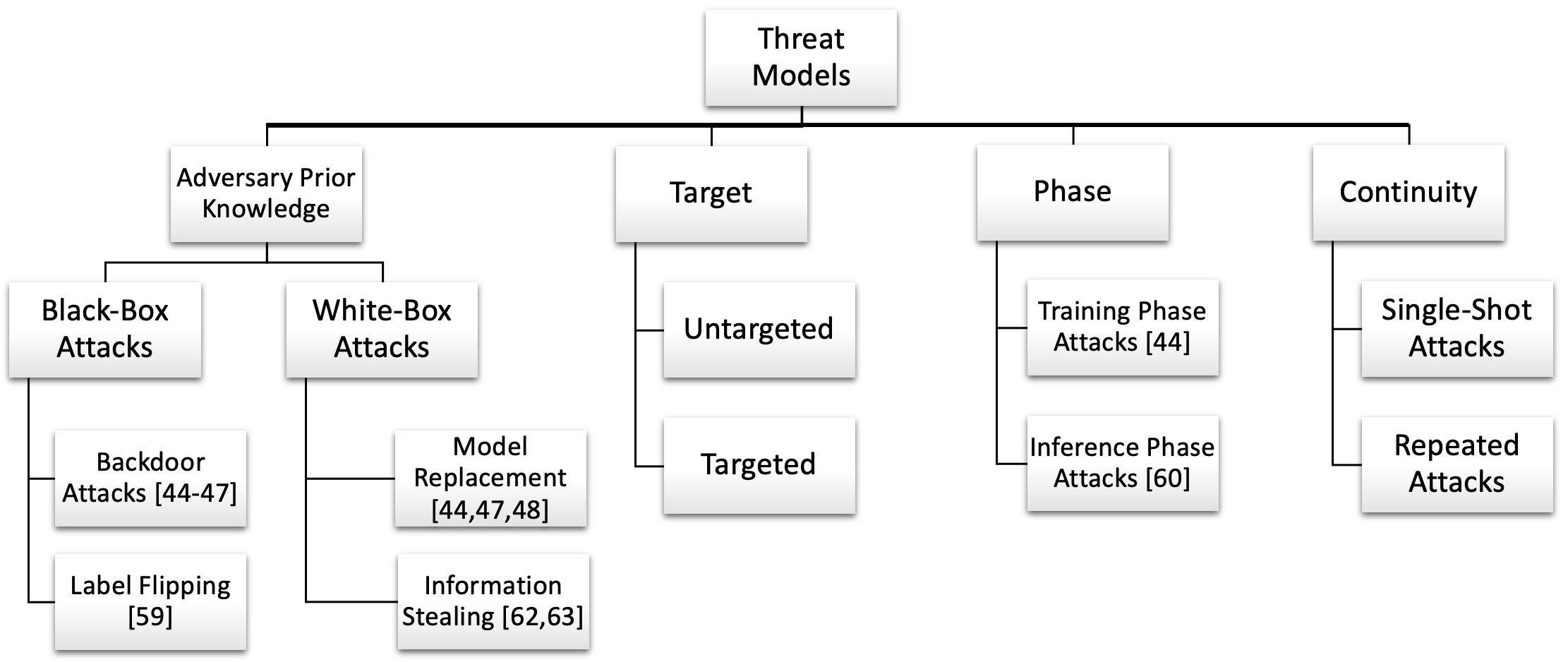}}
  \caption{Our taxonomy for threat models in decentralized deep learning systems.}
  \label{sun7}
\end{figure*}

Our taxonomy for threat models (Fig. \ref{sun7}) comprehensively demonstrates various attacking methodologies in decentralized deep learning systems.

Given the level of an adversary's prior knowledge on compromised clients, attacks on FL can be divided into white-box attacks and black-box attacks. For the black-box setting, the attacker has only access to a client’s local data set and the objective is to replace the dataset with compromised backdoor data. The typical black-box attacks include backdoor attacks and label flipping attacks \cite{labelflipping}. On the other hand, for the white-box setting, the attacker is considered to have control over both the local data and the model of a client. In this case, the attacker can send back any malicious model it prefers to the PS. One typical white-box attack is the model replacement. 

Moreover, depending on the attacker’s objective, an attack is either an untargeted attack that aims to reduce the accuracy of the FL model or a targeted attack that aims to compromise the FL model to output an adversary-desired label. Besides, according to the attacking timing, an attack can be mounted at either the training phase \cite{28} or the inference phase \cite{44}, where the training phase refers to the model training in FL and the inference phase refers to the application after attaining a converged model.

In addition, the continuity of an attack has also influence on the attacking performance, where a single-shot attack usually involves one malicious participant who aims to inject a long-lasting malicious trojan into the model by mounting the attack in a single round of training and a repeated attack usually involves one or more malicious participants with a high possibility to be mounted in multiple rounds of training.

We offer an overview of the most effective attacks on FL in the following several sections, covering backdoor attacks, model replacement, and information stealing.

\subsubsection{Backdoor Attacks}
\label{Section 3.2.2}
\begin{figure}
\centerline{\includegraphics[width=\linewidth, trim=1 1 1 1,clip]{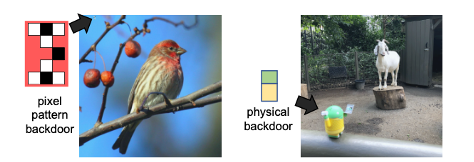}}
  \caption{Samples of various types of trojan backdoors. \cite{29}}
  \label{sun8}
\end{figure}

The goal of a backdoor attack is to corrupt other clients' model performance on specific sub-tasks. Given an attacker has only access to a client’s local data (a black-box attack), a trojan backdoor \cite{28, 29} corrupts a subset of local training data by adding a trojan pattern to the data and relabeling them as the target class (Fig. \ref{sun8}). Besides, Lin et al. \cite{30} adopted the composition of existing benign features and objects in a scene as the trigger. It leverages a mixer to generate mixed and poisonous samples, and then trains the local model on these samples as well as original benign data. Furthermore, semantic backdoors cause a model to produce an attacker-chosen output on unmodified images. For example, Wang et al. demonstrated an edge-case backdoor that targets prediction sub-tasks which are unlikely to be found in the training or test data sets, but are however natural \cite{31}. To conduct the attack, they trained the local model based on a mix of the backdoors and benign training data with a carefully chosen ratio. The result shows that this attack can bypass simple norm-based defense algorithms such as the norm bounding \cite{40}. 

\subsubsection{Model Replacement}
\label{Section 3.2.3}
The model replacement is one type of white-box attack \cite{28}, through replacing the global model with a malicious model. As aforementioned, in FedAvg, the PS updates the global model by performing a weighted average of all local trained models. A model replacement attack aims to submit a malicious model update $w_t^m=\frac{n_k}{n_{adv}}(w_t^{adv}-w_t) + w_t$ instead of $w_t^{adv}$, where $w_t^{adv}$ denotes a poisoned local model based on the aforementioned methods such as backdoor attacks and $n_{adv}$ denotes the number of samples owned by the adversary. Given the attack is usually mounted after the global model converges when additional local model training will not improve the global model and its loss settles within an error range around the optimum, each honest client$i$ then will obtain an updated local model $w_t^i$ approximately equal to the current global model $w_t$. $w_t^i-w_t\approx 0$. Equation (\ref{eq2}) is the mathematical proof of the model replacement attack. 

\begin{equation}
\label{eq2}
\begin{gathered}
w_{t+1}= w_t + \sum_{i \in k}\frac{n_i}{n_k} (w_t^i  - w_t)\\
= w_t+\frac{n_{1}}{n_k}(w_t^1 - w_t)+..+  \frac{n_{adv}}{n_k} (w_t^m  - w_t)\\
\approx w_t+\frac{n_{adv}}{n_k} (w_t^m  - w_t)\\
= w_t+\frac{n_{adv}}{n_k}(\frac{n_k}{n_{adv}}(w_t^{adv}-w_t) + w_t - w_t)\\
=w_t^{adv}
\end{gathered}
\end{equation}

The combination of semantic backdoors and model replacement formulates a long-lasting and invisible attack on the systems. For instance, Bagdasaryan et al. \cite{28} demonstrated an attacking method of the constrain and scale on the CIFAR-10 dataset, aiming to poison the global model using a set of car images with certain features (racing strip, green color, and stripped background wall) as triggers. In detail, the constrain-and-scale method is defined as in (\ref{eq3}). They mounted a single-shot model replacement attack where one malicious participant was selected for a single round in FL. Then, by updating the model based on the model prediction accuracy on both main classes and backdoor classes, and an anomaly detection algorithm’s accuracy, it aims to achieve the desired malicious performance while bypassing the anomaly detection. Finally, it shows that such attacks can bypass anomaly detection and retains a high accuracy for many rounds after the single-shot attack.

\begin{equation}
\label{eq3}
L_{model} = \alpha L_{class} + (1 - \alpha)L_{ano}
\end{equation}

Where $L_{class}$ captures the accuracy on both the main and backdoor tasks, $L_{ano}$ represents the performance of an anomaly detection algorithm taken at the PS, and $\alpha$ controls the importance of evading anomaly detection.

\subsubsection{Information Stealing  }
\label{Section 3.2.4}
By leveraging generative adversarial networks (GANs)\cite{gan}, an adversary could reconstruct the training data of another client in FL by just downloading the global model \cite{45}. In GANs, there is a tradeoff between the discriminator and the generator, where the discriminator trains based on whether it succeeds in distinguishing between the adversarial samples drawn from the generator and real data from the targeted data class, and the generator trains based on whether it succeeds in fooling the discriminator. At each round of FL, the adversary replaces the discriminator of the implemented GANs with the latest global model from the parameter server. Then the generator of the GANs produces adversarial samples from Gaussian noise and updates itself based on the inference result from the discriminator and the label of the targeted data class. In this case, with the adversarial training, the generator of the adversary could produce crispier samples to train a local DL model using the fake samples of the targeted data class. Besides, malicious model parameters of the adversary are then transmitted to the victim through model aggregation. The compromised model parameters would lure the victim to expose more detail on its training data, due to the victim would need more effort in model training thus identifying between real data and fake data. Consequently, the model update of the victim would allow the adversary to generate crispier and crispier adversarial samples that reveal the raw training data.

In addition, Nasr et al. \cite{8835245} demonstrated a comprehensive analysis on white-box membership inference attacks, where only correlated information of local training data leaked from the model sharing. Different from the aforementioned reconstruction attack where the objective is to reconstruct the raw training data of a victim, this kind of attack aims to infer whether a specific data sample was used in the victim’s local model training. 

\subsubsection{Defense Models}
\label{Section 3.2.5}
Defense strategies against the threat models in FL to date can mainly be separated into two categories, i.e., robust aggregation and anomaly detection. For the robust aggregation, instead of employing the random client selection policy in FedAvg, other selection approaches have been proposed against underlying malicious local updates, such as Krum\cite{33}, Trimmed mean\cite{34}, and so on. Krum selects a single local update from $m$ local models that is similar to other models as the global model based on pairwise Euclidean distances between local updates. In detail, for each model, it computes the sum of distances between a local model and its closest $m-c-2$ local models, where $m$ is the total number of clients, and $c$ is the assumed maximum number of compromised clients. On the other hand, trimmed mean sorts all local updates at each round, i.e., $w_{1j}$, $w_{2j}$, ···, $w_{mj}$, where $w_{ij}$  represents the $j$th round's model of the $i$th client. Then by removing the largest and smallest $\beta$ of them, the mean of the remaining $m - 2\beta$ models is employed as the result of the $j$th round’s global model. Moreover, another important strategy of robust aggregation is the Differential Privacy (DP), which limits the influence of a malicious update on model aggregation by adding a small fraction of Gaussian noise to the parameters of a local update. In particular, the cloud-side DP where the noise is added directly to the aggregated global model bounds the success of attacks such as the information stealing \cite{31, 37}. The client-side DP where the noise is added to each client's local update aims to achieve the optimized tradeoff between defense efficiency and model performance on main tasks of FL \cite{38}.  

Furthermore, norm bounding and anomaly detection are technologies adopted to drop malicious updates. In the norm bounding, the norm of a local update is a projected positive vector, such as the length of the model parameter vector. Since the malicious updates based on backdoor attacks and model replacement attacks of an adversary are likely to produce model parameters with large norms compared with other honest clients, an efficient way is to drop the updates whose norm is above a certain threshold \cite{40}. Likewise, anomaly detection in FL is usually based on comparing the similarity among local updates. For example, Cao et al. \cite{39} presented a Euclidean distance-based malicious local model detection. They demonstrated that if a local model had a distance under a certain constrain with more than half of the local models, it would be probably benign. Tolpegin et al. \cite{41} proposed a PCA-based defense against label flipping attacks. They observed and plotted standardized parameters of local updates to separate the benign and malicious ones. Additionally, Zhao et al. \cite{42} presented a poisoning defense method using generative adversarial networks. By reconstructing data from a local update and feeding the generated data to each of the clients’ models, they aimed to specify the label with the most occurrences as the true label for each input. Finally, the local updates were divided into the benign cluster and the malicious cluster by evaluating the prediction accuracy on the generated data using the obtained labels.

\section{Concluding Remarks}
\label{sec4}
In multi-access edge computing, decentralized deep learning (DDL) is considered to facilitate privacy-preserving knowledge acquisition from enormous various types of edge data. This survey provides an overview of DDL from two novel perspectives of communication efficiency and trustworthiness, offering state-of-the-art technologies to tackle challenges in leveraging DDL for social practices. 

Federated learning as a classical solution to data privacy in centralized learning, aims to leverage local model training for collective machine learning among multiple clients. Whereas, real-life challenges such as edge heterogeneity and adversarial attacks have greatly limited the capability and scalability of this technology. Given the capability limitation of an edge device, the convergence of a complicated model is costly and time-consuming. A more compatible architecture appears to be split learning, which brings the gap between a centralized computing resource and decentralized data sources. Besides, data heterogeneity is a common problem when applying real-world data, to this end, a more adaptive client selection policy could benefit the fast convergence of FL. Moreover, the topic of trustworthiness in DDL has also been attracting an explosive growth of interest in recent years. We summarized the latest threat models in DDL based on various criteria and provided our novel taxonomy. Finally, we discussed some of the most promising defense strategies against such threats on FL. In addition, there are still other important topics not covered in this survey, including mitigating algorithmic bias in DDL \cite{56, 62} and incentive mechanism for mobile device participation \cite{19, 68}.   

The current rapid advancement and broad application of deep learning in today’s society necessitates building trust in such emerging technology. The privacy-preserving DDL offers practical solutions to future large-scale multi-access edge computing. The breadth of papers surveyed suggests that DDL is being intensively studied, especially in terms of privacy protection, edge heterogeneity, and adversarial attacks and defenses. Furthermore, the future trends of DDL put weight on topics such as efficient resource allocation, asynchronous communication, and fully decentralized frameworks.

\section*{Acknowledgment}
The authors would like to thank the anonymous reviewers for helpful comments.

\bibliographystyle{IEEEtran}
\bibliography{TAI}

\begin{IEEEbiography}[{\includegraphics[width=1in,height=1.25in,clip,keepaspectratio]{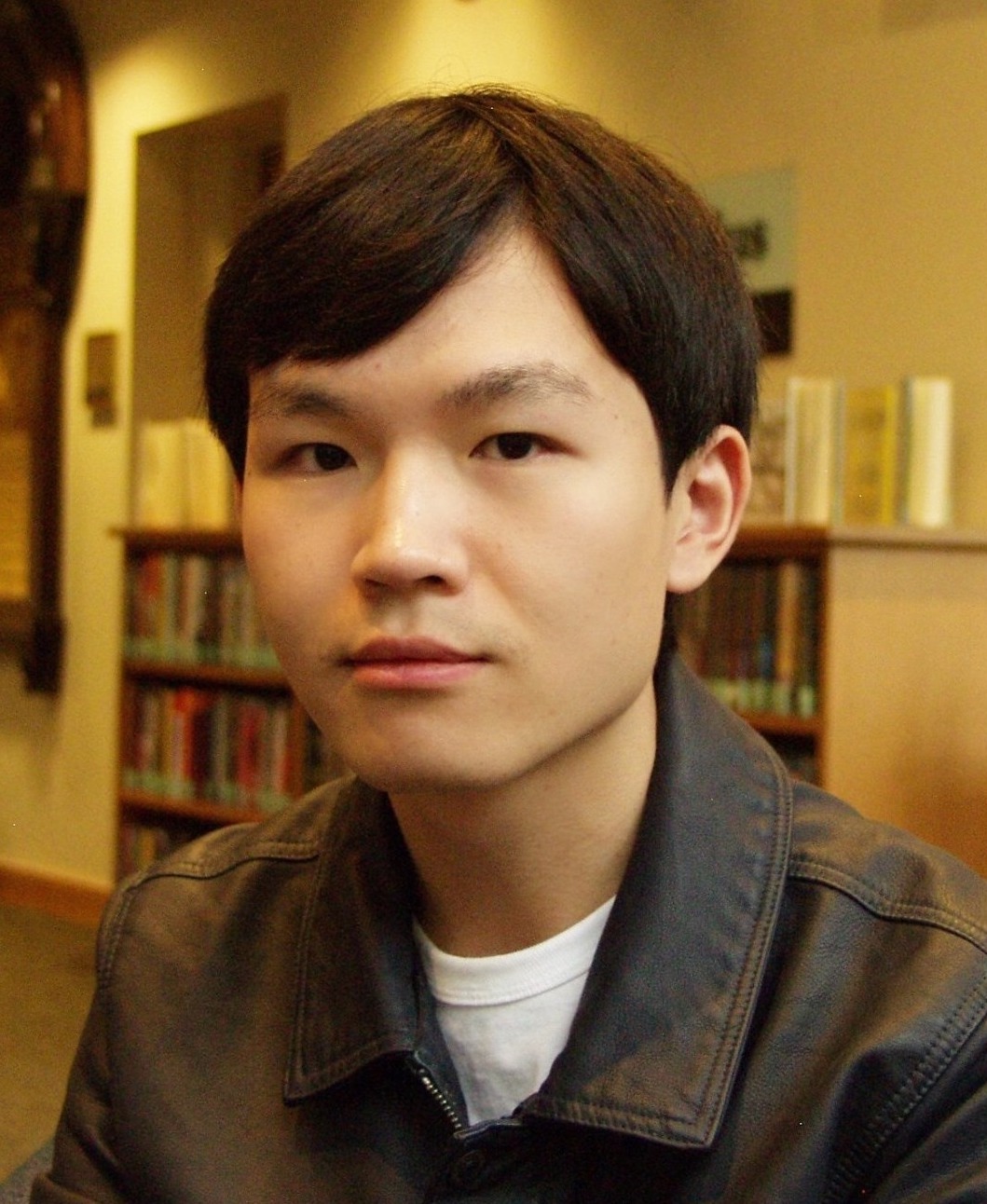}}]{Yuwei Sun}{\space}(M'20) is a Ph.D.’s student in the Graduate School of Information Science and Technology at the University of Tokyo. He received B.E. in Computer Science and Technology in 2018 from North China Electric Power University and M.E. in Information and Communication Engineering with honors in 2021 from the University of Tokyo. In 2020, he was the fellow of the Advanced Study Program (ASP) at the Massachusetts Institute of Technology. He has been working with the Campus Computing Centre at the United Nations University Centre on Cybersecurity since 2019. He is a member of the AI Security and Privacy Team at the RIKEN Center for Advanced Intelligence Project working on trustworthy AI, and a research fellow at Japan Society for the Promotion of Science (JSPS).  
\end{IEEEbiography}

\begin{IEEEbiography}[{\includegraphics[width=1in,height=1.25in,clip,keepaspectratio]{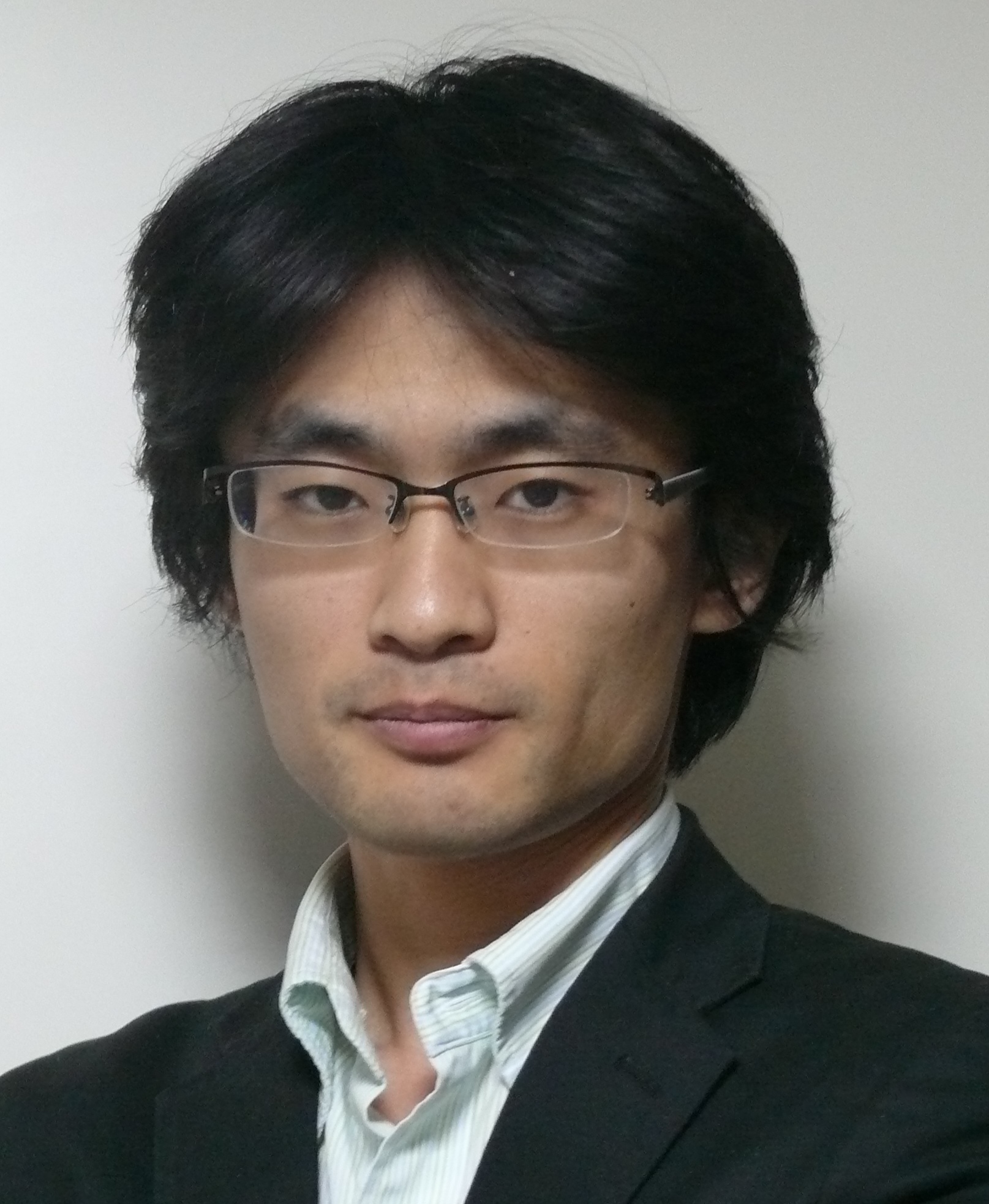}}]{Hideya Ochiai}{\space}(M'10) is an associate professor of the University of Tokyo, Japan. He received B.E. in 2006, M.E. in 2008, and Ph.D. in 2011 from the same university. His research interests have been sensor networking, delay tolerant networking, and building automation systems, IoT protocols, and cyber-security. He is involved in the standardization of facility information access protocol in IEEE1888, ISO/IEC, and ASHRAE.
\end{IEEEbiography}

\begin{IEEEbiography}[{\includegraphics[width=1in,height=1.25in,clip,keepaspectratio]{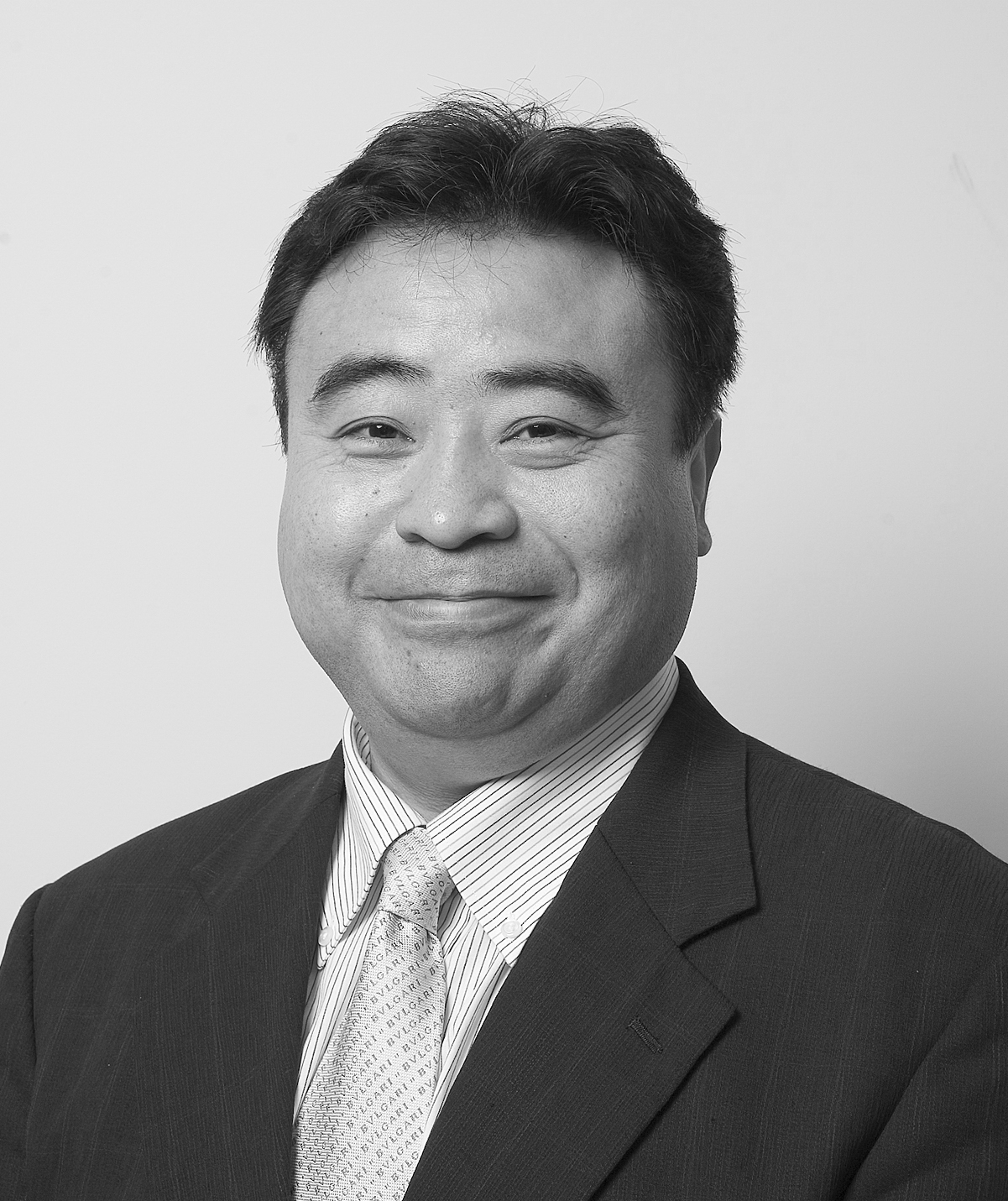}}]{Hiroshi Esaki}{\space}(M'08) received Ph.D. from the University of Tokyo, Japan, in 1998. In 1987, he joined Research and Development Center, Toshiba Corporation. From 1990 to 1991, he was at Applied Research Laboratory of Bell-core Inc., New Jersey, as a residential researcher. From 1994 to 1996, he was at Center for Telecommunication Research of Columbia University in New York. From 1998, he has been serving as a professor at the University of Tokyo, and as a board member of WIDE Project. Currently, he is the executive director of IPv6 promotion council, vice president of JPNIC, IPv6 Forum Fellow, and director of WIDE Project.
\end{IEEEbiography}
\end{document}